\documentclass[conference]{IEEEtran}
\IEEEoverridecommandlockouts
\usepackage{cite}
\usepackage{amsmath,amssymb,amsfonts}
\usepackage{algorithm}
\usepackage{algpseudocode}
\usepackage{graphicx}
\usepackage{textcomp}
\usepackage{xcolor}
\usepackage{booktabs}
\usepackage{multirow}

\def\BibTeX{{\rm B\kern-.05em{\sc i\kern-.025em b}\kern-.08em
    T\kern-.1667em\lower.7ex\hbox{E}\kern-.125emX}}
\begin{document}

\bstctlcite{IEEEexample:BSTcontrol}

\title{ECG-Free Assessment of Cardiac Valve Events Using Seismocardiography \\
\thanks{*This work was supported by the National Science Foundation under Grant No. 2340020 and a SMART Business Act Grant through Mississippi Institutions of Higher Learning (Grant No. 2024-04).}
}

\author{\IEEEauthorblockN{Mohammad Muntasir Rahman}
		\IEEEauthorblockA{\textit{Dept of Ag. \& Biological Engineering} \\
			\textit{Mississippi State University}\\
			Mississippi State, MS 39762, USA \\
			mmr510@msstate.edu}
		\and
		\IEEEauthorblockN{Aysha Mann}
		\IEEEauthorblockA{\textit{Dept of Ag. \& Biological Engineering} \\
			\textit{Mississippi State University}\\
			Mississippi State, MS 39762, USA \\
			ajm1034@msstate.edu}
		\and
		\IEEEauthorblockN{Amirtahà Taebi}\thanks{Corresponding author: ataebi@abe.msstate.edu}
		\IEEEauthorblockA{\textit{Dept of Ag. \& Biological Engineering} \\
			\textit{Mississippi State University}\\
			Mississippi State, MS 39762, USA \\
			ataebi@abe.msstate.edu}		
}
	
\maketitle

\begin{abstract}
Seismocardiogram (SCG) signals can play a crucial role in remote cardiac monitoring, capturing important events such as aortic valve opening (AO) and mitral valve closure (MC). However, existing SCG methods for detecting AO and MC typically rely on electrocardiogram (ECG) data. In this study, we propose an innovative approach to identify AO and MC events in SCG signals without the need for ECG information. Our method utilized a template bank, which consists of signal templates extracted from SCG waveforms of 5 healthy subjects. These templates represent characteristic features of a heart cycle. When analyzing new, unseen SCG signals from another group of 6 healthy subjects, we employ these templates to accurately detect cardiac cycles and subsequently pinpoint AO and MC events. Our results demonstrate the effectiveness of the proposed template bank approach in achieving ECG-independent AO and MC detection, laying the groundwork for more convenient remote cardiovascular assessment.

\end{abstract}


\section{Introduction}
Cardiovascular diseases (CVDs) continue to be a major health threat worldwide, affecting millions and straining healthcare systems. In the United States alone, a quarter of the population falls victim to CVDs each year, highlighting the urgent need for better prevention and management methods \cite{tsao2022heart,roth2020global}. Early detection of cardiac issues is paramount for improving patient outcomes. Seismocardiography (SCG) is a non-invasive means of cardiac monitoring, capturing subtle vibrations generated by the mechanical activity of the beating heart \cite{taebi2019recent,taebi2017time, zanetti1991seismocardiography,rahman2023non,rahman2023reconstruction}. Unlike electrocardiography (ECG) which relies on electrical signals, SCG focuses on the mechanical aspects of cardiac function and can be integrated into wearable devices \cite{brunelli2009template} to provide complementary information to other common cardiac assessment modalities. The SCG signals can be used to determine key events, including aortic valve opening (AO) and mitral valve closure (MC) \cite{mann2024exploring}. AO corresponds to the moment when the aortic valve opens during systole, allowing blood to flow from the left ventricle into the aorta, while, MC marks the closure of the mitral valve at the end of diastole, preventing backflow of blood into the left atrium. These events are crucial for understanding cardiac dynamics and diagnosing some cardiovascular conditions \cite{inan2014ballistocardiography}.

While the combination of ECG and SCG signals provides precise timing information for AO and MC, relying solely on one of these modalities can lead to the development of more convenient and compact cardiac monitoring methods. Previous studies have explored ECG-free heartbeat detection in SCG signals using template matching \cite{centracchio2023ecg,parlato2023ecg,centracchio2024accurate}, but these approaches often rely on templates derived from the same SCG signals being analyzed. Our research aims to bridge this gap by developing an ECG-independent algorithm for AO and MC detection using SCG signals only, without relying on the templates derived from the same signals under analysis. We leverage a template bank, a collection of signal segments extracted from SCG recordings of a group of subjects, and utilize it to detect heartbeats in the SCG signals of a new group of subjects. These templates represent characteristic features of the heart cycle, allowing us to detect heartbeats even in the absence of ECG information. The idea is to create a collection of optimal templates that can be used to search for heart cycles in new SCG signals through template matching.

Our proposed method identifies the optimal template from the template bank for a given SCG signal, and then, detects the heart cycles based on this template via template matching. Template matching involves locating occurrences of a specific template within a larger signal. The primary aim is to pinpoint regions in the signal that closely resemble the template. This process entails searching for particular features or segments within a new signal to determine if any similarities or identical matches exist. Typically, this similarity assessment is quantified by evaluating normalized cross-correlation values across the signal. In the subsequent sections, we provide details of the methodology, template creation, and assessment of the effectiveness of the proposed method. 

\section{Materials and Methods}
\subsection{Participants}
The research study received approval from the Institutional Review Board (IRB) at Mississippi State University. Data were collected from a group of 11 human subjects, comprising 7 males and 4 females (age: 23.00 $\pm$ 4.60 years, BMI: 24.30 $\pm$ 4.37 kg/m²). All participants reported no history of CVDs.

\subsection{Experimental Setup}
\begin{figure}
	\centering
	\includegraphics[width=.75\columnwidth]{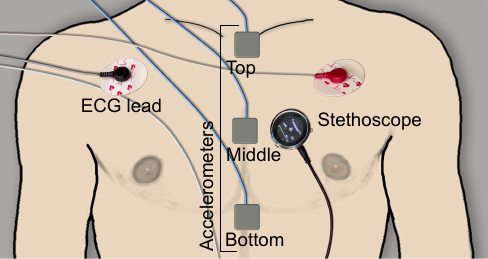}
	\caption{Data acquisition setup. The sensors include (a) a single-lead ECG, (b) three tri-axial accelerometers, and (c) an electronic stethoscope.}
	\label{fig:data_acquision}
\end{figure}

Subjects were instructed to lay in a supine position for sensor placement and data acquisition. Three tri-axial accelerometers (356A32, PCB Piezotronics, Depew, NY) were attached using double-sided tape to the manubrium, left sternal border near the fourth costal notch, and xiphoid process (Fig. \ref{fig:data_acquision}). We refer to these locations as ``top", ``middle", and ``bottom" for simplicity throughout the paper. These accelerometer outputs were amplified with a gain factor of 100 using a signal conditioner (model 482C, PCB Piezotronics, Depew, NY). Single-lead ECG signals were recorded, placing the electrodes on the participant following Einthoven's triangle configuration, under the left and right clavicle and on the right lower abdomen. Additionally, an electronic stethoscope (Thinklabs One, Thinklabs, Centennial, CO) was placed on the fourth left intercostal space, just to the right of the middle accelerometer. Participants were then instructed to remain still and breathe naturally throughout the data acquisition process. We monitored the signals in real time and ensured good signal quality before starting the recording. The data acquisition session began with a tap on the stethoscope and ended after approximately 2 minutes with another tap. All signals were recorded at a sampling frequency of 5000 Hz. In this study, the stethoscope served solely to identify the beginning and end of each measurement session through the tap sounds (Fig. \ref{fig:signal_trimming}).  

% Data acquisition

\subsection{Signal Preprocesssing}
\begin{figure}
	\centering
	\includegraphics[width=.65\columnwidth]{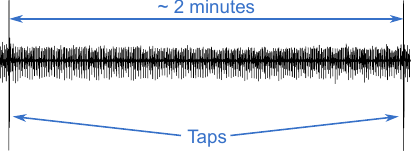}
	\caption{The spikes at the beginning and end of a raw PCG signal corresponding to the taps that were used to indicate the measurement session.}
	\label{fig:signal_trimming}
\end{figure}

MATLAB (R2022a, The MathWorks, Inc., Natick, MA) was used for data analysis. The electronic stethoscope output was used to identify the start and end points of the session. The stethoscope taps caused significant spikes in the signal (Fig. \ref{fig:signal_trimming}) which were used to trim the ECG and SCG signals. Digital filters were used to eliminate noise from both the ECG and SCG signals. For the ECG signals, we applied a bandpass filter with cutoff frequencies of 0.5 and 40 Hz, targeting the typical range of heart electrical activity. For all SCG signals, we used a 1–30 Hz bandpass filter. These cutoff frequencies for the SCG signals were chosen to eliminate the low-frequency respiration movement while preserving the dominant frequencies of the heart vibrations.

% Signal Trimming

\subsection{Identifying Heartbeats Using Template Matching}
Our primary goal is to detect AO and MC events in the SCG signal without relying on ECG data. To achieve this goal, we first leverage a template bank to identify the cardiac cycles on the SCG signal ECG-independently.

\subsubsection{Template Bank Creation}  
\begin{figure}
	\centering
	\includegraphics[width=0.75\columnwidth]{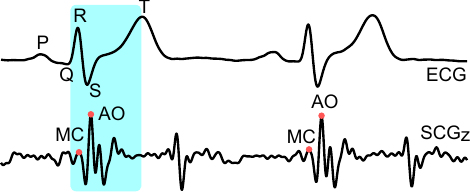}
	\caption{The SCG templates were created using a systolic window from the ECG Q wave to the T peak, representing the time for both ventricular depolarization and repolarization.}
	\label{fig:template_creation}
\end{figure}

\begin{figure}
	\centering
	\includegraphics[width=\columnwidth]{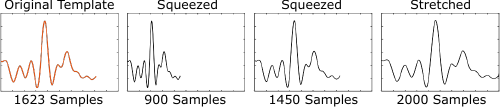}
	\caption{Example of template augmentation. The original template is resampled at 900:20:2000 sample points. Templates resampled at 900, 1450, and 2000 sample points are shown here.}
	\label{fig:template_resampling}
\end{figure}

To create a template, we define a systolic window using the ECG Q wave and T peak on the SCG signals (Fig. \ref{fig:template_creation}). Our template bank includes one template from each of subjects 7 to 11 (S07-S11, consisting of 2 female subjects) for every measurement location, i.e., the top, middle, and bottom sensor locations in Fig. \ref{fig:data_acquision}. This is done because SCG signals vary not only between individuals but also across chest locations. To account for different heart rates, we augment the templates by ``squeezing" and ``stretching" them. A subject with a lower heart rate would have a more stretched systolic phase while one with a higher heart rate would have a more squeezed heartbeat. Therefore, the template bank is augmented by resampling each template to a minimum of 900 and a maximum of 2000 sample points, with intervals of 20 points (Fig. \ref{fig:template_resampling}) resulting in a total of 855 templates in the template bank. This approach ensures that our template bank represents heartbeats at various heart rates ($\sim$45-10 bpm assuming that QT interval is 30\% of a cardiac cycle).

% Template Creation

% Template Resampling (stretch and squeeze)

\subsubsection{Optimal Template Selection}
The goal is to find the most suitable template in the template bank for detecting the SCG heartbeats. The pseudocode for this process is shown in Algorithm \ref{alg:find-best-template}. The algorithm begins by estimating the minimum number of heartbeats expected in the signal, assuming a minimum heart rate of 40 bpm. Subsequently, it iterates through each template in the bank, calculating the normalized cross-correlation (NCC) score between the template and the SCG signal. The maximum NCC score is then identified. If the maximum score exceeds the current best-performing score and the number of detected peaks meets the minimum required heartbeats, the best-performing score is updated. Finally, after all iterations, the algorithm returns the template that achieved the highest performance. Given the total number of templates in the bank, it is computationally expensive to use the entire 2-minute SCG data for each subject to find the best template. To address this, we only utilize the first 10 seconds of a given SCG signal to identify the optimal template from the template bank using the aforementioned algorithm. Once the best template is identified, it will be used to detect SCG heartbeats over the entire 2-minute SCG signal.

% Algorithm to find best template
\begin{algorithm}[b]
	\small
	\caption{Find Best Template from Template Bank}\label{alg:find-best-template}
	\begin{algorithmic}[1]
		\Function{FindBestTemplate}{$newSCG, templateBank, f_s$}		
		
		\State Calculate the length of $newSCG$ in seconds: \newline
		\hspace*{2.7em} $time\_length \gets length(newSCG)/fs$
		
		\State Minimum heart rate (bpm) to consider: \newline
		\hspace*{2.7em} $min\_hb \gets 40$ 
		\State Minimum number of heartbeats based on $min\_hb$: \newline 
		\hspace*{2.7em} $min\_n\_hb \gets round(time\_length \times min\_hb/60)$
		
		\State Initialize $bestScore \gets -\infty$
		\State Initialize $bestTemplate \gets [\hspace{.15cm}]$
		
		\For{$i = 1$ to $length(templateBank)$}
		\State Read template: $t \gets templateBank[i]$
		
		\State Calculate NCC: $score \gets ncc(template, newSCG)$		
		
		\State Find max Score: $maxScore \gets \max(score)$		
		
		\State Define a threshold: $k \gets maxScore - 0.25$
		
		\State Find peaks: \newline 
			\hspace*{4em} $locks \gets findpeaks(score, MinPeakHeight = k,$ \newline
			\hspace*{13em} $MinPeakDistance = fs/2)$
		
		\If{$maxScore > bestScore$ and \newline
			\hspace*{7em} $length(locks) >  min\_n\_hb$}		
		\State Update: $bestScore \gets maxScore$
		\State Update: $bestTemplate \gets template$
		\EndIf
		\EndFor
		
		\State \Return $bestTemplate$
		\EndFunction
	\end{algorithmic}
\end{algorithm}

\subsubsection{Heartbeat Detection}
The NCC between the best template and the SCG signal was calculated by sliding the template along the entire signal. The underlying assumption is that the template exhibits quasi-periodicity within the signal, with peaks corresponding to the heartbeats. By analyzing the NCC, we identify the highest peaks, pinpointing the locations of maximum similarity with the template. These identified locations correspond to the heartbeats in the signal. To detect these peaks, we utilized the ``\textit{findpeaks}" function in MATLAB, specifying a minimum peak height of maximum NCC score minus 0.25 and a minimum peak distance of half of the sampling frequency. The choice of subtracting 0.25 from the maximum NCC score was based on empirical tuning to ensure robustness against noise and signal variations. 

% Hearbeats detection
\begin{figure}[b]
	\centering
	\includegraphics[width=\columnwidth]{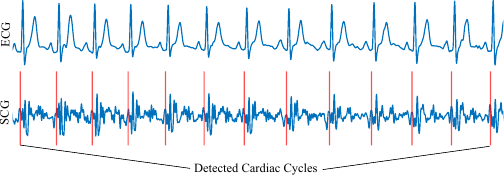}
	\caption{Heartbeats detected by the proposed algorithm for a new subject (S02). The bank does not include any templates from this subject.}
	\label{fig:heartbeats}
\end{figure}

\subsubsection{AO and MC Events Detection}
We used a similar technique described in \cite{mann2024exploring} by defining a search window starting from the beginning of each detected cardiac cycle to identify AO and MC. Typically, within this window, AO corresponds to the second prominent peak, while MC is the peak immediately preceding the AO. We evaluated the performance of our algorithm for AO and MC detection on the SCG signals recorded from the middle location. To do this, we compared the algorithm’s output with manually determined indices of AO and MC points. These points were selected by a researcher blind to the algorithm’s results. The researcher received training based on relevant SCG literature and manually identified the AO and MC points on the SCG signals for each subject using ECG information. To manage data collection efficiently, we developed a MATLAB code that stored the time indices of these fiducial points upon the researcher’s selection. We then calculate the precision~=~TP$\mathbin{/}$(TP+FP), recall~=~TP$\mathbin{/}$(TP+FN), and F1-score~=~2$\times$(Precision$\times$Recall)$\mathbin{/}$(Precision+Recall), where, TP represents the number of true positives, i.e., correctly detected AO and MC points. FP stands for false positives, indicating points identified by the algorithm that were not present in the manually labeled AO and MC points. Finally, FN refers to the false negatives, representing manually annotated AO and MC events that the algorithm missed.

\section{Results and Discussion}
\subsection{SCG Heartbeat Detection}
Fig. \ref{fig:heartbeats} shows an example of detected heartbeats for subject 2 (S02) using the proposed approach. S02 is a new subject and the template bank does not include any templates created from the SCG signals of this subject. The algorithm had a similar performance on other subjects. These results indicate that our approach allows accurate detection of cardiac cycles without relying on simultaneous ECG signals.

\subsection{AO and MC Detection Performance Evaluation}
To evaluate the performance of our AO and MC detection algorithm, we calculated precision, recall, and F1-score. Precision reflects the proportion of correctly detected AO and MC points out of all detected points. Recall indicates the proportion of correctly detected AO and MC points out of all manually selected AO and MC points. Finally, the F1-score provides a harmonic mean of precision and recall, offering a balanced view of detection performance.

The results are summarized in Table \ref{tab:10_sec_training}. The table includes precision, recall, and F1 scores for both unseen subjects (S01-S06) and subjects used in the design of the template bank (S07-S11).
Subjects within the template bank achieved an average precision of 97.96\% and 98.29\% for AO and MC detection, respectively. The average recall for these subjects was 92.85\% and 93.14\% for AO and MC, respectively. The average F1-score, which provides a balanced view of precision and recall was 95.20\% for AO and 95.59\% for MC. These figures demonstrate that our algorithm successfully detects SCG heartbeats in a 2-minute SCG signal for each subject, relying solely on a template bank containing just one template from that specific SCG signal.

For the unseen subjects, i.e., those subjects that no templates were created from their SCG signals and were not included in the template bank, the average precision was slightly lower at 96.23\% and 95.55\% for AO and MC detection, respectively. The average recall was also slightly lower at 81.94\% and 81.60\% for AO and MC, respectively. Finally, the average F1-scores for unseen subjects were 86.28\% and 85.71\% for AO and MC, respectively. S06 exhibited a lower precision, recall, and F1-score than other unseen subjects due to the best template failing to match the signal's systolic cycles accurately, resulting in some diastolic parts of the signal being mistakenly detected. This issue arises from person-to-person variability in the SCG signal. When evaluating the performance of the AO and MC detection algorithm on all subjects, the average precision was 97.02\% and 96.79\%, the average recall was 86.90\% and 86.85\%, and the average F1-score was 90.34\% and 90.20\% for AO and MC detection, respectively. These results suggest that our ECG-free method effectively detected SCG heartbeats and consequently identified important cardiac valve events including the timing of AO and MC.

%% all signal from my experiment with 10 second signal training
\begin{table}[t]
	\centering
	\small
	\caption{Performance of the Proposed ECG-Independent Algorithm in detecting the AO and MC points.}
	\label{tab:10_sec_training}
	\setlength{\tabcolsep}{3pt}
	\begin{tabular}{llcrrrrrr}
		\toprule
		&& \multirow{2}{*}{\textbf{Subjects}} & \multicolumn{2}{c}{\textbf{Precision (\%)}} & \multicolumn{2}{c}{\textbf{Recall (\%)}} & \multicolumn{2}{c}{\textbf{F1 Score (\%)}} \\
		\cmidrule(lr){4-5} \cmidrule(lr){6-7} \cmidrule(lr){8-9}		
		&&  & AO & MC & AO & MC & AO & MC \\
		\cmidrule(llr){1-3} \cmidrule(lr){4-5} \cmidrule(lr){6-7} \cmidrule(lr){8-9}
		
		\multirow{6}{*}{\rotatebox[origin=c]{90}{Unseen}}& \multirow{6}{*}{\rotatebox[origin=c]{90}{Subjects}}& S01 & 100.00 &  99.62 & 100.00 &  99.61 & 100.00 &  99.42  \\
		&& S02 &  98.75 &  98.75 &  91.86 &  91.86 &  95.18 &  95.18 \\
		&& S03 &  95.24 &  94.29 &	 86.96 &  86.09 &  90.91 &  89.57 \\
		&& S04 & 100.00 &  99.34 &  94.41 &  93.79 &  97.12 &  96.18 \\
		&& S05 &  97.26 &  97.94 &  92.81 &  93.46 &  94.98 &  95.98 \\
		&& S06 &  86.11 &  83.33 &  25.62 &  24.79 &  39.49 &  37.93 \\	
		\midrule		
		\multirow{5}{*}{\rotatebox[origin=c]{90}{Template}} & \multirow{5}{*}{\rotatebox[origin=c]{90}{Bank}} &S07 &  99.31 & 100.00 &  99.31 & 100.00 &  99.31 & 100.00 \\ 
		&& S08 &  94.89 & 	94.89 &  94.20 &  94.20 &  94.54 &  94.54 \\			
		&& S09 &  97.14 & 	97.14 &  90.07 &  90.07 &  93.47 &  93.47 \\
		&& S10 &  99.42 & 	99.42 &  99.42 &  99.42 &  99.42 &  99.42 \\
		&& S11 &  99.05 & 100.00 &  81.25 &  82.03 &  89.27 &  90.52 \\	
		
		\bottomrule
	\end{tabular}
\end{table}

\subsection{Future Directions}
Future work will focus on refining the criteria for selecting the best template from the template bank. This involves developing advanced algorithms to evaluate template similarity more effectively and incorporating machine learning techniques to improve the accuracy and reliability of AO and MC detection. Additionally, we plan to explore adaptive template matching methods that can dynamically adjust to individual variations in heart rate and signal morphology. These enhancements aim to further improve the proposed ECG-independent approach in real-world settings.

\section{Conclusion}
We presented an ECG-independent method for detecting AO and MC using SCG. By creating a template bank from systolic SCG templates collected from a group of five healthy subjects, we identified heartbeats using template matching without relying on ECG data. The template bank was made to account for the variability in SCG signals due to both inter- and intra-subject differences as well as heart rate variations. Our approach establishes a robust foundation for creating a versatile template bank capable of detecting cardiac cycles in new SCG signals, enabling accurate AO and MC detection even in the absence of ECG data. This method offers a promising avenue for developing more convenient cardiovascular assessment methods based on SCG.

\bibliographystyle{IEEEtran}	
\bibliography{references}	

\end{document}